\documentclass[pra,twocolumn,superscriptaddress,groupedaddress,showpacs,amsmath,amssymb,floatfix]{revtex4}
\usepackage{graphicx}
\usepackage{dcolumn}
\usepackage{bm}
\def\btt#1{\texttt{\@backslashchar#1}}
\DeclareRobustCommand\bblash{\btt{\@backslashchar}}

\def\>{\rangle}
\def\<{\langle}

\begin{document}

\title{One-way quantum computation with four-dimensional photonic qudits}

\author{Jaewoo Joo}%\email{jaewoo.joo@imperial.ac.uk}
\affiliation{Blackett Laboratory, Imperial College London, Prince
Consort Road, London, SW7 2BW, UK}
\author{Peter L. Knight}%\email{p.knight@imperial.ac.uk}
\affiliation{Blackett Laboratory, Imperial College London, Prince
Consort Road, London, SW7 2BW, UK}
\author{Jeremy L. O'Brien}
\affiliation{Centre for Quantum Photonics, H. H. Wills Physics
Laboratory \& Department of Electrical and Electronic Engineering,
University of Bristol, Merchant Venturers Building, Woodland
Road, Bristol, BS8 1UB, UK}
\author{Terry Rudolph} %\email{t.rudolph@imperial.ac.uk}
\affiliation{Blackett Laboratory, Imperial College London, Prince
Consort Road, London, SW7 2BW, UK}
\affiliation{Institute for
Mathematical Sciences, Imperial College London, London, SW7 2BW,
UK}

\date{\today}

\begin{abstract}

We consider the possibility of performing linear optical quantum
computation making use of extra photonic degrees of freedom. In
particular we focus on the case where we use photons as quadbits,
4-dimensional photonic qudits. The basic 2-quadbit cluster state
is a hyper-entangled state across polarization and two spatial
mode degrees of freedom. We examine the non-deterministic methods
whereby such states can be created from single photons and/or
Bell pairs, and then give some mechanisms for performing
higher-dimensional fusion gates.

\end{abstract}
% 03.67.Lx, 03.67.-a, 03.65.Ud
\maketitle
\section{Introduction}
\label{Intro} Optical quantum computation is a strong candidate
for a scalable quantum computer. Photons have low decoherence
rates, and high fidelity optical components are readily
available. In this article we focus on the linear optical quantum
computation (LOQC) paradigm, for which the resource overheads of
the original LOQC proposal \cite{Knill01} have been greatly
reduced by making use \cite{Nielsen04,
Browne05,Kok,kieling06,kieling07,Gross,Rohde,
Dawson06,Dawson062,Gilbert,Zhang,varnava} of the one-way quantum
computation model \cite{Briegel01,Raussendorf01}.

Significant hurdles to practical LOQC remain, however. At present
the primary obstacle is a deterministic source of photons. Much
progress has been made along these lines
\cite{Walther05,everyone}, but it is clear that there is still a
long way to go. Particularly exciting is the possibility of
creating ``on-demand'' entangled pairs of photons \cite{toshiba,
gershoni}, which obviate the need for initially creating such
entangled pairs from single photons \cite{pan}. The
investigations of this paper are based around an assumption that
at some time in the near future efficient deterministic sources
of either single photons or entangled photon pairs will become
available.

It is not always obvious how to compare the resource requirements
of various different proposals for implementing LOQC within the
cluster state paradigm (e.g. how many single photon sources,
memory units and feedforward steps is an entangled pair source
``worth''?). Since the primary difficulties for LOQC relate to
sources and detectors, it is clear that schemes which reduce the
number of photons actually used in an implementation are
desirable \footnote{In general, within the circuit model, the
results of \cite{Muthukrishnan} suggest that  quantum computation
with quadbits can be expected to result in a space saving of at
least $O(\log_2 d)$, and a time saving of at least $O(\log_2
d)^2$. The extent to which such savings translate into the
cluster state model are largely unexplored - they will depend on
optimal decompositions of qudit cluster circuits for general
two-qudit unitary operations. Such optimal decompositions are not
completely characterized for even the qubit case yet.}. A
travelling photonic wavepacket is in principle a multi-mode
creature, and thus can be treated as a $d$-dimensional quantum
system (a ``qu$d$it''). There is a $d$-dimensional version of
cluster state computing \cite{Zhou03, BillHall}, and one purpose
of this paper is to explore procedures whereby such
$d$-dimensional clusters can be created. The second motive is to
examine some basic ``initial state'' resource tradeoffs, such as:
``how many Bell pairs does it take to make a hyper-entangled
state''.

For concreteness we focus on the \emph{quadbit} case -
specifically, we treat a single photon as a four-dimensional
quantum system; using the two polarization states of two
different spatial modes to encode the four levels.

\section{Quadbit cluster states}
\label{Sec:01}
\subsection{General quadbit cluster states}
\label{Sec:0101}

In this section we review the features of quadbit cluster states
we shall make use of - a pedagogical overview of the
higher-dimensional cluster state computing can be found in
\cite{BillHall}.

We label the computational basis states $\{|\bar0 \rangle ,|\bar1 \rangle,
|\bar2 \rangle  ,|\bar3 \rangle \}$ (use of the overbar is to prevent
confusion with 0 and 1 photon Fock states). In terms of these we can
define the quadbit version of a Hadamard rotation, which rotates the
computational basis state $|\bar i\>$ to $|+_i\>$ ($i=0,1,2,3$), where
\begin{eqnarray}
\label{five:eq03} | +_{i} \rangle &=& {1\over 2} ( |\bar0 \rangle
+ {\rm e}^{{\rm i} {i \pi \over 2}} |\bar1 \rangle + {\rm
e}^{{\rm i} \,i \pi}|\bar2 \rangle + {\rm e}^{{\rm i} {3 i \pi
\over 2}} |\bar3 \rangle),
\end{eqnarray}

A 2-quadbit cluster state $|QdC_2\>$ is then given by the
superposition
\[|QdC_2\>={1 \over 2}\sum_{i=0}^3|\bar i\>|+_i\>,\] which should be compared with
the equivalent 2-qubit cluster state
$|C_2\>=(|0\>|+\>+|1\>|-\>)/\sqrt{2}$. In the case of qubits a
two-qubit (non-destructive) parity gate operation would fuse
\cite{Browne05} two 2-qubit clusters into the state
$|C_3\>=(|+\>|00\>|+\>+|-\>|11\>|-\>)/\sqrt{2}$, and repeated such
fusion operations allows for the growth of arbitrary cluster
states (the redundant encoding of the central qubit is easily
removed by a measurement in the $|\pm\>$ basis, yielding the
3-qubit cluster state as claimed). Similarly, in the quadbit
cluster case arbitrary quadbit clusters can be grown using a
quadbit fusion operation. Applied to two 2-quadbit clusters such
a fusion would achieve the state
$|QdC_3\>=\sum_{i=0}^3|+_i\>|\bar i \bar i\>|+_i\>/2$.

\subsection{Optical quadbit cluster states}
\label{Sec:0102}

We define a quadbit single photon quantum state in two
polarization/spatial modes as follows:
\begin{eqnarray}
\label{five:eq01} |\bar0 \rangle &\equiv& |H \rangle_1 \,
,~|\bar1 \rangle \equiv |V \rangle_1 \, , ~ |\bar2 \rangle \equiv
|H \rangle_{2} \, ,~|\bar3 \rangle \equiv |V \rangle_{2} \, ,~~~~~
\end{eqnarray}
where $H (V)$ denotes horizontal (vertical) polarization,  and
the subindex $1 (2)$ denotes spatial mode $k_1$($k_2$).

Consider now a so-called hyper-entangled state (HES)
\cite{kwiat01}, which is a two-photon  state entangled in both
polarization and spatial modes.  Two-photon HES's can be
generated by spontaneous parametric down-conversion \cite{kwiat}.
As with generation of single photons, such a mechanism of HES
production is not scalable. In Section \ref{Sec:0201}, we will
consider scalable production of HES's given deterministic single
photon sources or entangled pairs. It is possible to represent an
HES  as product of Bell states, with a virtual tensor product
structure between the spatial and polarization modes, for example:
\begin{eqnarray}
\label{five:eq05}
|\Phi^{+}_{\rm HES} \rangle &=& {1 \over 2} ( | H \rangle | H
\rangle + |V \rangle |V \rangle)\otimes(|1\rangle|3\rangle +
|2\rangle|4\rangle).~~~~~
\end{eqnarray}
(We will always use the $\otimes$ symbol to refer to this virtual
tensor product of spatial modes and polarizations). Using the
identification in Eq.~(\ref{five:eq01}), we see that
$|\Phi^{+}_{\rm HES} \rangle$ is equal to
\begin{eqnarray*}
\label{five:eq06} |\Phi^{+}_{\rm HES} \rangle &=& {1\over 2}
(|H\>_1|H\>_3+|V\>_1|V\>_3+|H\>_2|H\>_4+|V\>_2|V\>_4)
\\&=& {1\over 2} (
|\bar0 \rangle |\bar0 \rangle + |\bar1 \rangle |\bar1 \rangle +
|\bar2 \rangle |\bar2 \rangle + |\bar3 \rangle |\bar3 \rangle).
\end{eqnarray*}
As any single mode unitary operation can be implemented with
linear optics \cite{reck}, a simple circuit can be constructed
which rotates the quadbit in modes 3 and 4 to yield the optical
2-quadbit cluster state $|QdC_2\>$ defined above.
%\begin{eqnarray}
%\label{five:eq07} |{\rm QdC_2}\rangle &=& {1\over 2} ( |\bar0
%\rangle | +_{0} \rangle + |\bar1 \rangle | +_{1} \rangle + |\bar2
%\rangle | +_{2} \rangle + |\bar3 \rangle | +_{3} \rangle),~~~~~~
%\end{eqnarray}
%which is a two-quadbit cluster state by definition.

Consider attempting to fuse two 2-quadbit clusters, the first in
modes (1,2;3,4) (as in Eq.(\ref{five:eq05})), the second in modes
(5,6;7,8). The procedure required to fuse the quadbit in spatial
modes 1,2 with that in 5,6 is a gate which (when successful)
performs a projective measurement of the form:
\[
|HH\>_{1\,5}\<HH|+|VV\>_{1\,5}\<VV|+|HH\>_{2\,6}\<HH|+|VV\>_{2\,6}\<VV|.
\]
That is, a successful measurement should reveal ``the photons
were in corresponding spatial modes with the same polarization'',
but should not reveal in which spatial modes and with what
polarization. In section \ref{Sec:03} we will show that such a
fusion is possible, although we have only found methods of doing
it that make use of ancillary systems, and for which the success
probability strongly depends on the nature of the ancillas
available.

%In the following section, two deterministic and probabilistic methods
%are presented to generate a hyper-entangled state from different
%initial states and to build Type2-like fusion gate and
%multi-quadbit cluster states from two HESs.

\section{Generation of quadbit cluster states}
\label{Sec:02}

Before discussing possible fusion mechanisms, we turn to
examining some
 ``initial state resource tradeoffs''. This is because, as in the case
 of single photons, parametric downconversion is not a suitable source
 for scalable LOQC. Therefore we may well need to generate deterministic
 HES's from a deterministic source of either single photons, Bell pairs or
 GHZ states. Whether the constructions we give are optimal (or even close
 to being so) we cannot determine. Thus the procedures we present should be
 seen as simply giving upper bounds on the resources required. Also,
 because the most efficient fusion gate we will present for quadbit clusters destroys
 the photons involved (much like Type-II fusion for qubits) we will need to look
 at mechanisms for generating an initial resource of 3 and 4 quadbit
 cluster states.

Basic notation for the figures, and a brief outline of the
operation of the fundamental optical components is set out in
Appendix A.

\subsection{General procedure for HES generation}
\label{Sec:0201}

\begin{center}
\begin{figure}[h]
\resizebox{!}{6.5cm} {\includegraphics{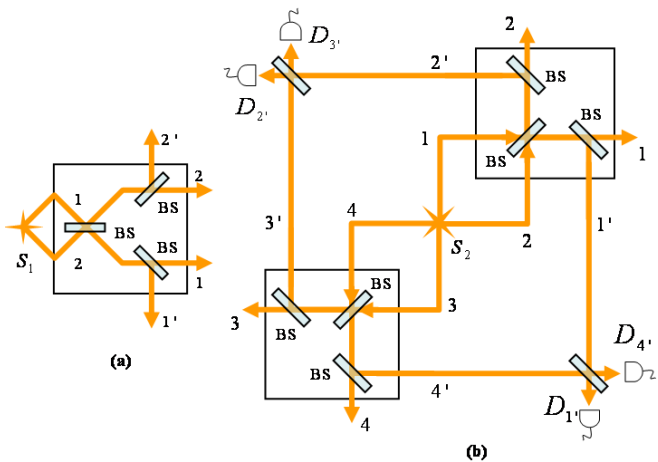} } \caption{(color
online) \label{fig:J2} (a) Circuit $J_1$ (b) Circuit $J_2$ for a
hyper-entangled state from four entangled photons}
\end{figure}
\end{center}

The general circuit we present (Fig.\ref{fig:J2}) is built from
two copies of a sub-circuit we label $J_1$, and we first explain
the operation of this circuit.

The circuit $J_1$ consists of three beam splitters (BSs) with two
vacuum inputs. Consider the case where a Bell state $(|H\>_1
|H\>_2+|V\>_1|V\>_2)/\sqrt{2}$ is input into $J_1$.  The first BS
creates a bunched two-photon state in modes 1 and 2, and then two
vacuum inputs are applied from modes $1'$ and $2'$ with two
regular BSs. After the circuit $J_1$, the state of two photons in
mode 1, $1'$, $2$, and $2'$ is equal to
\begin{eqnarray}
\label{six:eq02} &&|M \rangle_{121'2'} = {1 \over 4} ( | H \rangle
| H \rangle + |V \rangle |V \rangle) \nonumber \\ &&~~~~~~~\otimes
 \sum_{j=1}^{2}\big[{\rm
e}^{{\rm i}j\pi}(|j\rangle|j\rangle +|j'\rangle|j'\rangle +
\sqrt{2} |j\rangle|j'\rangle ) \big].~~~~~~~
\end{eqnarray}
It is a combination of four states of bunched photon pairs in
a spatial mode ($|j\rangle|j\rangle$ and $|j'\rangle|j'\rangle$)
and two anti-bunched states in two different spatial
modes ($|j\rangle|j'\rangle$).

We turn now to the full circuit $J_2$ depicted in Figure
\ref{fig:J2} (b). At the centre of the circuit is a source $S_2$
which can be either single photons, Bell pairs or a 4-photon GHZ
state. This source is then fed into two copies of the $J_1$ gate,
the outputs of which impinge on 50:50 beam splitters as shown. It
is easiest to begin with the case that the source consists of two
Bell pairs.

The initial state of the two Bell pairs $|{\Phi^{+}} \rangle_{1\,2}
|{\Phi^{+}} \rangle_{3\,4}$ is
\begin{eqnarray}
\label{six:eq04}  && {1 \over 2} ( | H \rangle_1 | H \rangle_2 +
|V \rangle_1 |V\rangle_2 )( | H \rangle_3 | H \rangle_4 + |V
\rangle_3 |V\rangle_4 ).
\end{eqnarray}
According to Eq. (\ref{six:eq02}), the state after the two $J_1$
circuits is equal to
\begin{eqnarray}
\label{six:eq045} && \hspace{-1cm} |M \rangle_{1\,2\,1'\,2'} |M
\rangle_{3\,4\,3'\,4'}\nonumber \\
&=& {1 \over 16}( | H \rangle | H \rangle + |V \rangle |V
\rangle)( | H \rangle | H \rangle + |V \rangle |V \rangle)
\nonumber \\&& ~~~\otimes \sum_{j=1}^{2}\big[{\rm e}^{{\rm
i}j\pi}(|j\rangle|j\rangle +|j'\rangle|j'\rangle + \sqrt{2}
|j\rangle|j'\rangle ) \big] \nonumber \\&& ~~~~~~
\sum_{k=3}^{4}\big[{\rm e}^{{\rm i}k\pi}(|k\rangle|k\rangle
+|k'\rangle|k'\rangle + \sqrt{2} |k\rangle|k'\rangle ) \big].~~~
\end{eqnarray}

At the end of the $J_1$ circuits, two BSs are applied in modes
$1',4'$ and $2',3'$, after which detectors are located. Successful
operation occurs when two identically polarized photons are
detected in modes $1', 4'$ or $2', 3'$ respectively, and the
success probability of the detection pattern is 1/16. To see how
this works, note that it is the components of the state in Eq.
(\ref{six:eq045}) which consist of two bunched photons
($|j'\rangle|j'\rangle$ and $|k'\rangle|k'\rangle$) that can
yield successful detection : the anti-bunched photonic states
($|j\rangle|j'\rangle$ and $|k\rangle|k'\rangle$) result in
destructive interference. For example, if we detect two
horizontal photons in modes $1'$ and $4'$ but nothing in modes
$2'$ and $3'$, the outcome state is
\begin{eqnarray}
\label{six:eq06} | \psi'_{\rm HES} \rangle &=& {1 \over 2
\sqrt{2}} ( | H
\rangle | H \rangle + |V \rangle |V \rangle) \nonumber \\
&& \otimes (|1\rangle|1\rangle - |2\rangle|2\rangle +
|3\rangle|3\rangle - |4\rangle|4\rangle).~~~~~
\end{eqnarray}
This state is, up to a linear optical transformation (in this case
two BSs in mode 1 and 2 and mode 3 and 4), a hyper-entangled
state.

It is interesting to note that the failure outcomes can still
yield photons in useful states. In particular the failure outcome
where only the vacuum is detected leaves all the photons still in
two Bell pairs ; this occurs with probability 1/16, and obviously
the gate can then simply be repeated. This suggests the overall
success probability is essentially 1/8. Some of the detection
patterns, while not yielding an HES do still leave two of the
photons in Bell pair, which could be recycled.

We are also able to use for the source a four-qubit GHZ state of
the form $(|HHHH\>_{1234}+|VVVV\>_{1234})/\sqrt{2}$ rather than
two Bell pairs; this yields a higher success probability. This
also has the advantage that in this case we need not assume the
four detectors are polarization sensitive : they need only count
numbers of photons at the output of the primed modes. Upon
successful detection, when two photons are detected in any two
spatial modes, the state in modes 1 to 4 becomes a HES with a
success probability 3/16, which is higher than the case of two
Bell pairs. Interestingly, no photon detection yields a 4-photon
entangled state such as $(|\Phi^{+}\>_{12}
|\Phi^{+}\>_{34}+|\Psi^{+}\>_{12} |\Psi^{+}\>_{34})/\sqrt{2}$.

Finally, if we wish to create a HES ballistically from single
photons, then we can replace the two Bell pairs input at the
source $S_2$ by two copies of the circuit for generating a Bell
pair from 4 single photons (Figure \ref{NewS01} in Appendix
\ref{Append03}). In this case we find that the success probability
is $1/16^3$.

\subsection{Generating larger quadbit cluster states}
\label{Sec:0202}
\subsubsection{3 quadbit cluster state}
\label{Sec:020201}
\begin{figure}[h]
\centering
\includegraphics[width= 9cm]{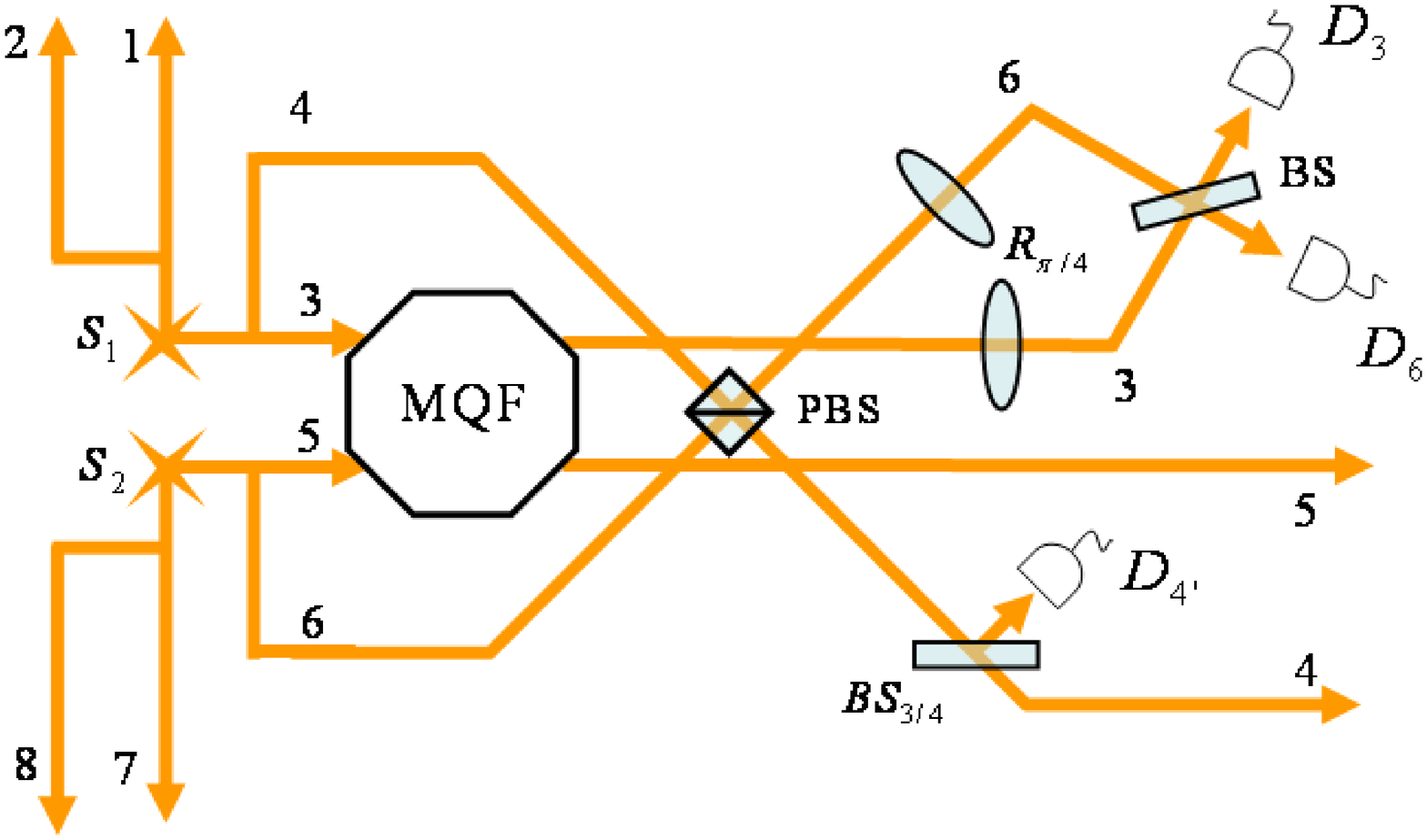} \vspace{-1cm}
\caption{ \label{MQF02} (color online) Circuit $K_1$ for a 3
quadbit cluster state from two HESs}
\end{figure}

To create a 3 quadbit cluster state, we use the ``modified quantum
filter'' (MQF) scheme we present in Appendix \ref{Append0202}.
This circuit implements a parity gate between the input photons
in a manner which does not destroy the input photons when it is
successful, and moreover is unaffected by situations wherein one
of the input modes is empty.

Our circuit for generating a 3 quadbit cluster from two HES's is
depicted in Figure \ref{MQF02}. $S_1$ and $S_2$ are sources of
initial HESs each in $|\Phi^{+}_{\rm HES} \rangle$. Note that
there is \emph{one} photon spread across spatial modes (3,4) and
one photon spread across spatial modes (5,6) - the circuit is a
two-photon gate, and only one photon will be detected - this is
reminiscent of fusing together two Bell pairs by Type-I fusion to
create a 3 qubit GHZ (cluster) state,  and in fact this gate does
act as a Type-I fusion gates for quadbits. After a successful
operation in modes 3 and 5 of the MQF, the outcome state is equal
to
\begin{eqnarray}
\label{seven:eq02} && {\sqrt{2} \over 6} \big[ |H\rangle_1
|H\rangle_3 |H\rangle_5 |H\rangle_7 + |V\rangle_1 |V\rangle_3
|V\rangle_5 |V\rangle_7 \nonumber \\ &&~+ 2 (|H\rangle_2
|H\rangle_4 + |V\rangle_2 |V\rangle_4) (|H\rangle_6
|H\rangle_8 + |V\rangle_6 |V\rangle_8 ) \big], \nonumber \\
\end{eqnarray}
(the measurement operator for operation of the MQF's is presented
in Eq. (\ref{QF04}) in Section \ref{Append0202}). Note that in Eq.
(\ref{seven:eq02}), the first two terms contain a photon in mode
3 and the other terms also have a photon in mode 6 (these are the
modes which will be detected). After a polarizing beam splitter
(PBS) between modes 4 and 6, two $R_{\pi/4}$s, and a BS in mode 3
and 6, detection of a photon in either mode 3 or 6 results in
successful gate operation. The outcome state results from only
four terms in Eq. (\ref{seven:eq02}), such as $|H\rangle_1
|H\rangle_3 |H\rangle_5 |H\rangle_7 $, $|V\rangle_1 |V\rangle_3
|V\rangle_5 |V\rangle_7$, $|H\rangle_2 |H\rangle_4 |H\rangle_6
|H\rangle_8$, and $|V\rangle_2 |V\rangle_4 |V\rangle_6
|V\rangle_8$. The extra beam splitter (${\rm BS}_{3/4}$) with
vacuum input in mode 4 balances amplitudes in the final state.
For example, after a successful detection in the MQF, the
detection of a vertical photon in mode 3 and vacuum in modes 6
and $4'$ yields a final state
\begin{eqnarray}
\label{seven:eq03} |{ QdC'_3} \rangle &=& {1\over 2} \big(
|H\rangle_1 |H\rangle_5 |H\rangle_7 - |V\rangle_1 |V\rangle_5
|V\rangle_7 \nonumber \\ &&~~~+ |H\rangle_2 |H\rangle_4
|H\rangle_8 - |V\rangle_2
|V\rangle_4 |V\rangle_8 \big), \nonumber \\
&=& {1\over 2} \big( |\bar0 \rangle |\bar0 \rangle |\bar0 \rangle
- |\bar1 \rangle |\bar1 \rangle |\bar1 \rangle + |\bar2 \rangle
|\bar2 \rangle |\bar2 \rangle - |\bar3 \rangle |\bar3 \rangle
|\bar3 \rangle \big),\nonumber \\
\end{eqnarray}
where the set $\{|\bar0 \rangle, |\bar1 \rangle, |\bar2 \rangle,
|\bar3 \rangle\}$ is defined by $ \{|H\rangle_1$, $|V\rangle_1$,
$|H\rangle_2$, $|V\rangle_2\}$, $ \{|H\rangle_5$, $|V\rangle_5$,
$|H\rangle_4$, $|V\rangle_4\}$, and $ \{|H\rangle_7$,
$|V\rangle_7$, $|H\rangle_8$, $|V\rangle_8\}$.

When the generalized quadbit Hadamard operation and a phase shift
are employed on a vertical photon in mode 5 and 4, the outcome
state is equivalent to a 3-quadbit cluster state $|{
QdC_3}\rangle$ in Section \ref{Sec:0101}. Therefore, we obtain a
three-quadbit cluster state in modes 1,2,4,5,7, and 8 with success
probability 1/256.

\begin{figure}[h]
\centering
\includegraphics[width= 9cm]{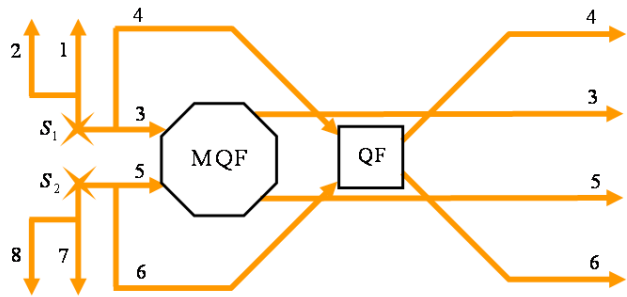} \vspace{-1.5cm}
\caption{ \label{7th} (color online) Circuit $K_2$ for a 4-quadbit
cluster state from 2 hyper-entangled pairs}
\end{figure}

\subsubsection{4 quadbit cluster state}
\label{Sec:020202} A slight modification of the circuit in the
previous subsection can easily build a 4-quadbit cluster state.
We start from the intermediate state in Eq. (\ref{seven:eq02})
(see Figure \ref{7th}). Because the state does not contain an
input of vacuum states in mode 4 and 6, the original QF can be
used (see Section \ref{Append0202}). When the original QF is
successfully applied in modes 4 and 6 to the outcome in Eq.
(\ref{seven:eq02}), the final state is equal to
\begin{eqnarray}
\label{seven:eq05} |{\rm QdC'_4}\rangle &=& {1\over 2} \big(
|H\rangle_1 |H\rangle_3 |H\rangle_5 |H\rangle_7 + |V\rangle_1
|V\rangle_3 |V\rangle_5 |V\rangle_7 \nonumber \\  && ~+
|H\rangle_2 |H\rangle_4 |H\rangle_6 |H\rangle_8 + |V\rangle_2
|V\rangle_4 |V\rangle_6 |V\rangle_8 \big).\nonumber \\
\end{eqnarray}
This is equivalent to
\begin{eqnarray}
\label{seven:eq055} |{\rm QdC_4}\rangle &=& {1\over 2}
\sum^{3}_{d=0} | +_{d} \rangle | \bar{d} \rangle | +_{d} \rangle
| +_{d} \rangle,
\end{eqnarray}
up to a local operation on the second photon. Note this is a
4-quadbit state of ``star'' form - i.e a central quadbit with
three leaves, and thus is useful for creating quadbit clusters
with nontrivial topology.

 From the resource point of view, two
hyper-entangled states and six single photons (four horizontal and
two vertical photons) are used to create such a 4 quadbit cluster
with success probability 1/1024.

\section{Fusing quadbit cluster states}
\label{Sec:03}

\begin{center}
\begin{figure}[h]
\centering
\includegraphics[width=11cm]{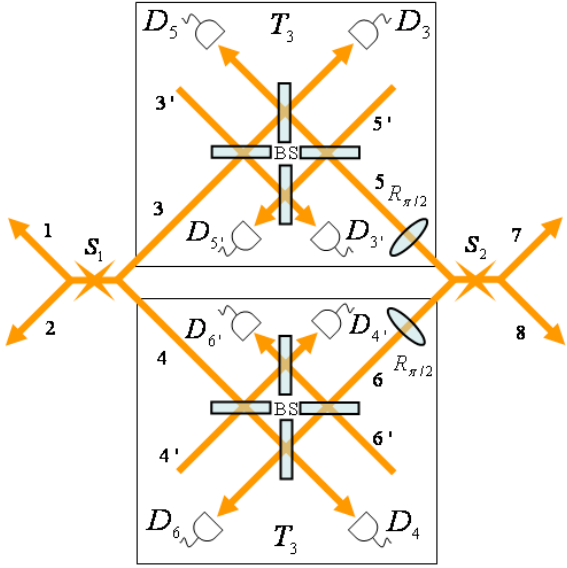} \caption{ \label{8th}
(color online) Circuit $K_3$ of a Type2-like fusion gate on two
hyper-entangled pairs}
\end{figure}
\end{center}

In order to perform optical quadbit one-way quantum computation,
we require a procedure for building large multi-quadbit cluster
states. The Type-I style gate (of section \ref{Sec:020201}) could
be used; however its success probability is very low.

In Figure \ref{8th}, we present a Type-II-like fusion gate
between two quadbit cluster states. The total circuit is
comprised of two sub-circuits we label $T_3$, consisting of two
four-port interferometers. The operation of the $T_3$ gate is
discussed in Appendix \ref{Append0201}. The basic effect of gate
$T_3$ is to destroy the spatial mode information carried by the
photons while leaving their polarization information in fact.

As shown in Figure \ref{8th}, the initial state is prepared in
$|\Phi^{+}_{\rm HES} \rangle_{1\,2\,3\,4} |\Phi^{+}_{\rm HES}
\rangle_{5\,6\,7\,8}$. What we desire of this gate is that when
it succeeds it tells us ``the photons were either in modes 3 and 5
or they were in modes 4 and 6, and their polarization was the
same''. However it should not reveal in which pair of spatial
modes they were, and with what polarization.

After two $R_{\pi/2}$s in mode 5 and 6, the
intermediate state is
\begin{eqnarray}
\label{seven:eq06} &&{1 \over 4} ( | H \rangle_1 | H \rangle_3 +
|V \rangle_1 |V\rangle_3 + | H \rangle_2 | H \rangle_4 + |V
\rangle_2 |V\rangle_4 )\nonumber \\
&&~~~( | V \rangle_5 | H \rangle_7 + |H \rangle_5 |V\rangle_7 + |
V \rangle_6 | H \rangle_8 + |H \rangle_6 |V\rangle_8 ).~~~~~~~~
\end{eqnarray}
Based on the discussion in Appendix \ref{Append0201}, if the upper
$T_3$ gate is implemented (without extra photons in modes $3'$,
$5'$) and a successful detection occurs (i.e. a single horizontal
and a single vertical photon are detected in two of the modes 3,
5, $3'$, and $5'$), it generates the Bell state in modes 1 and 7:
\begin{eqnarray}
\label{seven:eq07} (| H \rangle_{1} | H \rangle_{7} \pm |V
\rangle_{1} |V \rangle_{7})/\sqrt{2}.
\end{eqnarray}
Note that the parts of the input state with amplitude in modes 4,
6 are then wiped out.

On the other hand, if the lower $T_3$ gate detects one horizontal
and one vertical photon, originating from modes 4 and 6, it makes a Bell state in mode 2 and 8
\begin{eqnarray}
\label{seven:eq08} (| H \rangle_{2} | H \rangle_{8} \pm |V
\rangle_{2} |V \rangle_{8})/\sqrt{2}.
\end{eqnarray}
and amplitude for modes 3 and 5 is wiped out.

We essentially desire both of these $T_3$ gates to be able to
succeed simultaneously and indistinguishably. In order to attain
this, extra photons are injected into the spatial modes $3'$,
$4'$, $5'$, and $6'$. We will consider various possible initial
states for these ancillary photons. The basic idea is that
indistiguishable events occur if two photons in different
polarizations are detected in both the upper and lower $T_3$
gates simultaneously. These events can arise from either the
ancillary photons or the `actual inputs' - and our lack of
knowledge about which possibility occurs gives an amplitude for
both $T_3$ gates working. The success probability relies on the
input state of the extra two photons, and we discuss several
possibilities.

The first case is that two single photons are injected in mode
$3'$, $4'$, $5'$, and $6'$ in the state
\begin{eqnarray}
\label{seven:eq09} |{\rm Ex_1}\rangle &=& {1 \over 4} ( | H
\rangle_{3'} + |V \rangle_{3'} + | H \rangle_{4'} + |V
\rangle_{4'} )\,\nonumber \\&& ~~~~~( -| H \rangle_{5'} + |V
\rangle_{5'} -| H \rangle_{6'} + | V \rangle_{6'} ),~~~~~~~
\end{eqnarray}
where each photon is a superposed state in two spatial modes in
both polarizations.

When we detect two different polarized photons
in the upper $T_3$ gate and two different polarized photons in
the lower one, we do not know whether the four photons detected in
both $T_3$ gates come from hyper-entangled states or the extra input
photons. For example, the upper $T_3$ gate succeeds upon
detection of a horizontal photon in mode 3 and a vertical photon
 in mode 5. The photons could come from any two modes out of
modes 3, 5, $3'$, and $5'$. According to Eq. (\ref{seven:eq09}),
the detection works on various input states like $|H \rangle_{3}
|V \rangle_{5}$, $|V \rangle_{3} |H \rangle_{5}$ ,$|H \rangle_{3'}
|V \rangle_{5'}$, and $|V \rangle_{3'} |H \rangle_{5'}$. If the
detected photons were $|H \rangle_{3} |V \rangle_{5}$, $|V
\rangle_{3} |H \rangle_{5}$, the remaining state from Eq.
(\ref{seven:eq06}) is equal to the state in Eq.
(\ref{seven:eq07}). However, if the detected photons were $|H
\rangle_{3'} |V \rangle_{5'}$, and $|V \rangle_{3'} |H
\rangle_{5'}$, the lower circuit could be activated by $|H
\rangle_{4} |V \rangle_{6}$, $|V \rangle_{4} |H \rangle_{6}$ and
the remaining state equals Eq.(\ref{seven:eq08}). The same logic
can be applied  the other way around between the upper and lower
circuits. Thus, for the successful cases (two different polarized
photons detected in each $T_3$ gate), the final state is
equivalent to
\begin{eqnarray}
\label{seven:eq10} && {1 \over 2} ( | H \rangle_1 | H \rangle_7 +
| V \rangle_1 | V \rangle_7 + | H \rangle_2 | H \rangle_8 + | V
\rangle_2 | V \rangle_8 ) \,,~~~~~~~~
\end{eqnarray}
which is a superposition state of Eq. (\ref{seven:eq07}) and Eq.
(\ref{seven:eq08}). For this case, the total success probability
is 1/64.

We now consider injecting a Bell pair in mode $3'$, $4'$,
$5'$, and $6'$ instead of two single photons such as
\begin{eqnarray}
\label{seven:eq11} |{\rm Ex_2}\rangle &=& {1 \over 2\sqrt{2}}
\big[ ( | H \rangle_{3'} + |V \rangle_{3'})(-| H \rangle_{5'} + |V
\rangle_{5'}) \nonumber \\&&~~~~~+ (| H \rangle_{4'} + |V
\rangle_{4'})(-| H \rangle_{6'} + | V \rangle_{6'} ) \big].
\end{eqnarray}
it can readily be seen that the same indistinguishability of $T_3$
gate operations occurs - in this case with total success
probability 1/32.

Finally, the most efficient state to use is the ancillary input a
HES
\begin{eqnarray}
\label{seven:eq12} |{\rm Ex_3}\rangle &=& {1 \over 2} ( | H
\rangle_{3'}| V \rangle_{5'} +| V \rangle_{3'}| H \rangle_{5'}
\nonumber \\&&~~~~~+ | H \rangle_{4'}| V \rangle_{6'}+ | V
\rangle_{4'}| H \rangle_{6'}).~~
\end{eqnarray}
In this case the total success probability is 1/16.

Interestingly, even in some failure cases, we still have a chance
to have remanent entanglement between two photons in mode 1 (or
2) and 7 (or 8) of Figure \ref{8th}. Without the help of extra
photons in the primed modes, the success probability is 1/2 to
generate a Bell pair from two HESs through this circuit. If we use
two extra photons, the possibility of obtaining some entanglement
between 1 (or 2) and 7 (or 8) becomes higher than 1/2. This could
possibly be useful for some hybrid qubit/quadbit cluster states
computing schemes. For example, we imagine a modified qubit
cluster state possessing a HES at the edge and fuse two copies of
this state on the HES side in circuit $K_3$. With an extra HES in
the primed modes, a HES or a Bell pair is generated among mode 1,
2, 7, and 8 with overall probability 3/4.

We see therefore that we can use this Type2-like circuit to
create a Bell pair from HESs. As shown in Figure \ref{8th}, we
prepare two HESs with no extra photon. With probability 3/4 we
achieve a Bell state. Although this seems perverse - destroying
two HES's to create a Bell pair, it raises an interesting
possibility of attaching systems which have the form of a HES at
the end of the (qubit) cluster state. If we perform this fusion
gate on two such photons, it appears that we could fuse the larger
qubit cluster state with the probability 3/4.

\section{Summary of some resource tradeoffs}
\label{Sec:04}

\subsection{Difficulties of quantifying tradeoffs}
\label{Sec:0403}
%\subsubsection{Some basics of resource counting}

 A Bell pair can be created from 4 single photons with probability
1/4 (see Appendix \ref{Append03} for a proof - previously
published results \cite{pan} suggested the success probability
was 3/16). Such creation is \emph{ballistic} - the single photons
are fired in, and (up to some local linear transformation) the
desired Bell state is created 1 in 4 times. We could say that a
Bell pair is ``worth'' 16 single photons on average - this
indicates how much easier things will be if we have a
deterministic source of Bell pairs. Now a trivial extension of
this ballistic scheme can create a 3-photon GHZ state from 6
single photons  with probability 1/32, and can create
(ballistically) a 4-photon GHZ state with probability 1/128 (see
Table \ref{tab:MHES}). From this we might conclude a 3-photon GHZ
state is worth 96 photons. However we can also create a 3-photon
GHZ state by using a Type-I gate \cite{Browne05} and fusing two
Bell pairs. The Type-I gate succeeds with probability 1/2, and
each Bell pair is worth 16 photons, so this indicates the GHZ
state is only worth 64 single photons. The difference, of course,
is that with the latter technique we would have to store the Bell
pair, once created, in order for it to be available to combine
with the second Bell pair. While the ability to postselect on
successfully generated states (and then store them) lies at the
heart of why it is we can turn exponentially decreasing
probabilities into efficient methods for creating large entangled
states, such storage is likely to present practical problems. (It
is worth noting that the percolation techniques of
\cite{kieling06} ameliorate many of these issues).

Resource counting is made even messier by the following
observation: Sometimes we may require the use of an ancillary
entangled state within some larger ballistic circuit (a Bell pair
say). One may think that we could replace this Bell pair by 4
single photons (as in Fig. \ref{NewS01} in Appendix
\ref{Append03}) to obtain a ballistic single photon scheme, and
only take a hit of 1/4 in the overall success probability of the
larger circuit. However the ballistic scheme presumes the ideal
state is produced ``up to easily implementable linear optical
transformations'' - and it is generally a smaller set of
detection outcomes which yield the desired state for input into
the larger circuit.

The final feature that makes resource counting difficult is the
nature of failure outcomes: sometimes failed gates acting on
suitably large input states still leave some of the systems in
useful resource states. The potential for recycling (which also
requires quantum memory) often greatly complicates the question
of optimizing resource counting \cite{Gross}.
\begin{table}[b]
\centering
\begin{tabular}{|cccccc|c|c|}\hline
 &   &  & Resource &  & & Output & Probability \\ \cline{1-6}
 SP & BP & 3GHZ & 4GHZ & HES &
3QdC & &  \\
\hline 4 & &  & &  &
 & BP & 1/4 \\
6 & &  &  &  &
 & 3GHZ & 1/32 \\
8 & &  &  &  &
 & 4GHZ & 1/128 \\
 & 2 &  &  &  &
 & 3GHZ & 1/2 \\
 & 1 & 1 &  &  &
 & 4GHZ & 1/2 \\
\hline 8 & &  &  &  &
 & HES & 1/4096 \\
 & 2 &  &  &  &
 & HES & 1/16 \\
 &   &  & 1 &  &
 & HES & 3/16 \\
4 & &  & & 2 &  & 3QdC & 1/256 \\
6 & &  &  & 2 & & 4QdC & 1/1024 \\
6 & &  & & 1 & 1 & 4QdC &  1/256 \\
2 & &  &  &  & 2 & 4QdC & 1/64 \\
 & 1 &  &  &  & 2 & 4QdC & 1/32 \\
 & &  &  & 1 & 2 &  4QdC & 1/16 \\
\hline
\end{tabular}
\caption{ Resource costs for multi-quadbit cluster states (SP =
single photon, BP = Bell pair, HES = hyper-entangled state, and
QdC = quadbit cluster)} \label{tab:MHES}
\end{table}
\subsection{Resources for quadbit cluster states}
\label{Sec:0401}

As shown in Table \ref{tab:MHES}, A various combination of
resources can be used to create any desired state. Without an
entangled source ({i.e.} only single photon sources) one can
generate a Bell pair from 4 single photons, a 3-photon GHZ state
from 6 photons, and 4-photon GHZ state from 8 photons. However,
using entangled sources, the desired many-photon state can be
built with much higher probabilities.

In terms of quadbit cluster states, the counterpart of a Bell
pair for qubit is a HES. So, to build a HES requires a source
such as 8 single photons, two Bell pairs, or one 4-photon GHZ
state. Based on the circuit $J_2$ the optimal probability is
3/16, obtained when using a 4-photon GHZ state.

The bottom of the table shows various ways of building
multi-quadbit cluster states by proposed methods with the help of
extra photons. We only have one method to build 3 quadbit cluster
stated using circuit $K_1$, while several possible methods are
available to create a 4 quadbit cluster state through circuits
$K_2$ and $K_3$. For the 4 quadbit cluster state, the success
probability without a 3 quadbit cluster state is 1/1024 (using
circuit $K_2$).

\section{Conclusion}
\label{Conc} We have initiated the study of building higher
dimensional cluster states of photons. Although we have presented
several ``modules'' within our constructions that we expect to be
of generic use for LOQC using higher dimensional photonic states,
it is unclear to us whether the procedures we have outlined are
close to the best possible. If they are, then there seems to be
limited advantage in using higher dimensional cluster states
built up from single photons from a strict resource counting
perspective. It is possible, however, that in the future
deterministic sources of hyper-entanglement become available.

Very recently, qubit one-way quantum computation using a
hyper-entangled state (HES) has been demonstrated
\cite{new01,new02}. In these papers, a four-qubit cluster state is
created from a HES generated by a spontaneous parametric down
conversion. They assume that photon's polarization and its
spatial modes are defined as a qubit respectively and destroying
a single photon performs two single qubit measurements
simultaneously. Note that this is quite different to our
proposal, where we use a single photon as a higher-dimensional
quantum unit. An equivalence between these schemes arises for a
2-quadbit HES and a 4-qubit linear cluster state because
$2+2=2\cdot2$ \cite{new01,new02}.

\acknowledgements

We acknowledge useful discussions with Jens Eisert, David Gross
and Konrad Kieling and thank the support of the US Army Research
Office (W911NF-05-0397) and the UK EPSRC. This work was supported
in part by the UK Engineering and Physical Sciences Research
Council through their Quantum Information Processing
Interdisciplinary Research Centre, and by the European Union
through their networks SCALA and CONQUEST. J.J. was also
supported by the Overseas Research Student Award Program.

\appendix
\section{Basic optical tools}
\label{Append01}
\begin{center}
\begin{figure}[b]
\resizebox{!}{6.5cm} {\includegraphics{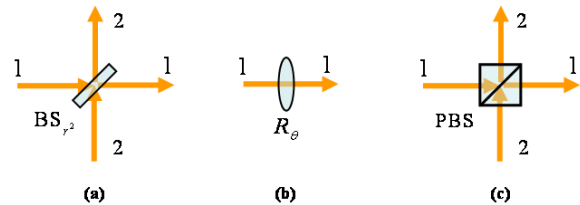}} \vspace{-2.5cm}
\caption{ \label{fig:Basic01} (color online) (a) Beam splitter
${\rm BS}_{r^2}$ (b) Polarization rotator $R_{\theta}$ (c)
Polarizing beam splitter (PBS) }
\end{figure}
\end{center}

We use the convention that a regular beam splitter (Fig.
\ref{fig:Basic01} (a)) which we denote ${\rm BS}_{r^2}$ acts upon
two spatial modes 1,2 according to
\begin{eqnarray}
\label{Chap601:eq01} |H (V) \rangle_1 &\rightarrow&  t |H (V)
\rangle_1 + r |H (V) \rangle_2 ,\nonumber \\ |H (V) \rangle_2
&\rightarrow& -r|H (V) \rangle_1 + t|H (V) \rangle_2 ,~~~
\end{eqnarray}
where $t^2+r^2=1$. A 50:50 BS ($t=r=1/\sqrt{2}$) acts according
to $|H (V) \rangle_1 \rightarrow (|H (V) \rangle_{1} + |H (V)
\rangle_{2}) / \sqrt{2}$ and $|H (V) \rangle_2 \rightarrow ( - |H
(V) \rangle_{1} + |H (V) \rangle_{2})/ \sqrt{2}$. On a single
photon this is a Hadamard gate in the spatial mode. However, when
two horizontal (vertical) photons are injected in modes 1 and 2
respectively, the total photonic state shows destructive
interference between modes 1 and 2 - the two photons bunch into a
single mode together. More explicitly, if two polarized photons
$|H \rangle_1 |H \rangle_2$ are injected onto a 50:50 BS, the
total state is equal to ${1 \over \sqrt{2}} |H \rangle |H \rangle
\otimes (- |1 \rangle |1 \rangle + |2 \rangle |2 \rangle)$ where
$|H^2 \rangle_{i}$ denote two horizontal photons in mode $i$.
This state could be called an entangled state in spatial mode. In
terms of the quadbit encoding Eq. \ref{five:eq01}, the regular BS
induces unitary rotations between state $|\bar 0\>$,$|\bar 2\>$
and between $|\bar 1\>$,$|\bar 3\>$.

In Fig. \ref{fig:Basic01} (b), a polarization rotator
($R_\theta$) in mode 1 is depicted - its action is taken to be
\begin{eqnarray}
\label{Chap602:eq01} |H \rangle_1 & \rightarrow& \cos \theta |H
\rangle_1 + \sin \theta |V \rangle_1  \, ,\nonumber \\ |V
\rangle_1 & \rightarrow& -\sin \theta |H \rangle_1
 + \cos \theta |V \rangle_1  ,
\end{eqnarray}
where angle $\theta$ is the amount of the rotating axis of linear
polarizer. A $\pi/4$ polarization rotator $R_{\pi/4}$ acts such
that $|H \rangle_1 \rightarrow |+ \rangle_1 = ( |H \rangle_{1} +
|V \rangle_{1} ) / \sqrt{2}$ and $|V \rangle_1 \rightarrow |-
\rangle_1 = (- |H \rangle_{1} + |V \rangle_{1}) / \sqrt{2}$. On a
single photon this is a Hadamard gate in polarization; on two
differently polarized photons in the same spatial mode $|H
\rangle_1 |V \rangle_1$ are transformed by a rotation $R_{\pi/4}$
into the state ${1 \over \sqrt{2}}( - |H \rangle |H \rangle + |V
\rangle |V \rangle ) \otimes |1 \rangle |1 \rangle$, again
because of destructive interference. This photonic state has
entanglement in polarization (although the two photons are located
in a single spatial mode together). In terms of the quadbit
encoding Eq. \ref{five:eq01}, the polarization rotator induces
unitary rotations between state $|\bar 0\>$,$|\bar 1\>$ and
between $|\bar 2\>$,$|\bar 3\>$.

In Fig.\ref{fig:Basic01}(c) a polarizing beam-splitter (PBS) is
depicted. It acts such that  a vertical photon is reflected into
a different spatial mode ($1 \rightarrow 2$ and $2 \rightarrow
1$), while a horizontal photon passes
 through a PBS. In terms of the quadbit
 encoding Eq. \ref{five:eq01}, the PBS induces a swap operation
 between $|\bar 1\>$ and $|\bar 3\>$.

\section{Advanced optical tools} \label{Append02}
\begin{center}
\begin{figure}[t]
\resizebox{!}{8.5cm} {\includegraphics{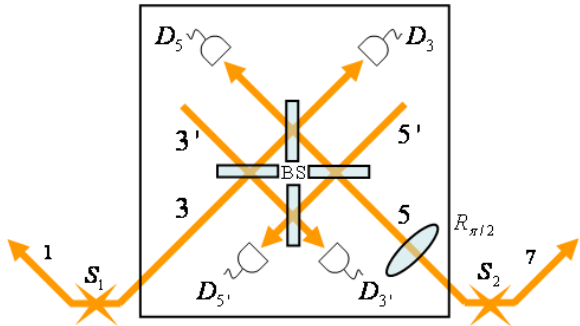} } \vspace{-1cm}
\caption{ \label{fig:T303} (color online) The Type3 fusion gate
$T_3$ creates indistinguishability of photons in mode 3 and 5}
\end{figure}
\end{center}
\subsection{Gate $T_3$}
\label{Append0201} Here we present a gate similar to a qubit
Type-II gate, which is used to create a Type-II-like gate for
quadbits. In Figure \ref{fig:T303}, we begin with two Bell-pairs
$|\psi^{+} \rangle_{1\,3} |\psi^{+} \rangle_{5\,7}$ and act the
gate on modes 3 and 5. The state after a $R_{\pi/2}$ in mode 5 is
equal to
\begin{eqnarray}
\label{T401} && {1 \over 2} ( | H \rangle_{1} | H \rangle_{3} + |V
\rangle_{1} |V \rangle_{3})( | V \rangle_{5} | H \rangle_{7} + |H
\rangle_{5} |V \rangle_{7}).~~~~~~~
\end{eqnarray}

A four-port interferometer is then introduced \cite{Zukowski}.
This interferometer essentially erases spatial mode information .
Its operation on a single photon input into any of the modes 3,
5, $3'$, $5'$ is:
\begin{eqnarray}
|P \rangle_3 & \rightarrow & \big( |P \rangle_3 + |P
\rangle_5 + |P \rangle_{3'} + |P \rangle_{5'}\big)/2 , \nonumber \\
|P \rangle_5 & \rightarrow & \big( - |P \rangle_3 + |P
\rangle_5 - |P \rangle_{3'} + |P \rangle_{5'}\big)/2, \nonumber \\
|P \rangle_{3'} & \rightarrow & \big( -|P \rangle_3 -
|P \rangle_5 + |P \rangle_{3'} + |P \rangle_{5'}\big)/2, \nonumber \\
|P \rangle_{5'} & \rightarrow & \big( |P \rangle_3 - |P \rangle_5
- |P \rangle_{3'} + |P \rangle_{5'}\big)/2, \label{4port01}
\end{eqnarray}
where $P$ denotes either horizontal or vertical polarization
($H,V$). In the case that inputs $3'$ and $5'$ are vacuum, as in
our example here, we need only consider here the first and second
transformations in Eq. (\ref{4port01}). Note, however, that the
possibility of extra input photons in these modes will be used in
Section \ref{Sec:03} to create indistinguishability in
polarization.

To continue with our example, when we detect a horizontal photon
in mode 3 and a vertical photon in mode 5, the original photons
in Eq.(\ref{T401}) could be either $|V \rangle_{3}$ $|H
\rangle_{5}$ or $|H \rangle_{3}$ $|V \rangle_{5}$ . When a
horizontal and a vertical photon are detected in any mode out of
3, 5, $3'$, and $5'$, the final state is
\begin{eqnarray}
\label{Type401} && {1 \over \sqrt{2}} ( | H \rangle_{1} | H
\rangle_{7} \pm |V \rangle_{1} |V \rangle_{7}).
\end{eqnarray}
The failure cases (the detection of two horizontal or two
vertical photons) yield a product state. That is, in the failure
cases, the detection of two horizontal (vertical) photons comes
only from $| H \rangle_{1} | V \rangle_{7}$ ($| V \rangle_{1} | H
\rangle_{7}$). The failure outcome gives us the polarization
information of the other photons.

The overall effect of the gate is to remove spatial mode
information in a way that does not destroy polarization
information. This gate $T_3$ has success probability 1/2 and it
destroys both input photons.

\subsection{Quantum filter (QF)}
\label{Append0202}
\begin{center}
\begin{figure}[b]
\resizebox{!}{12cm} {\includegraphics{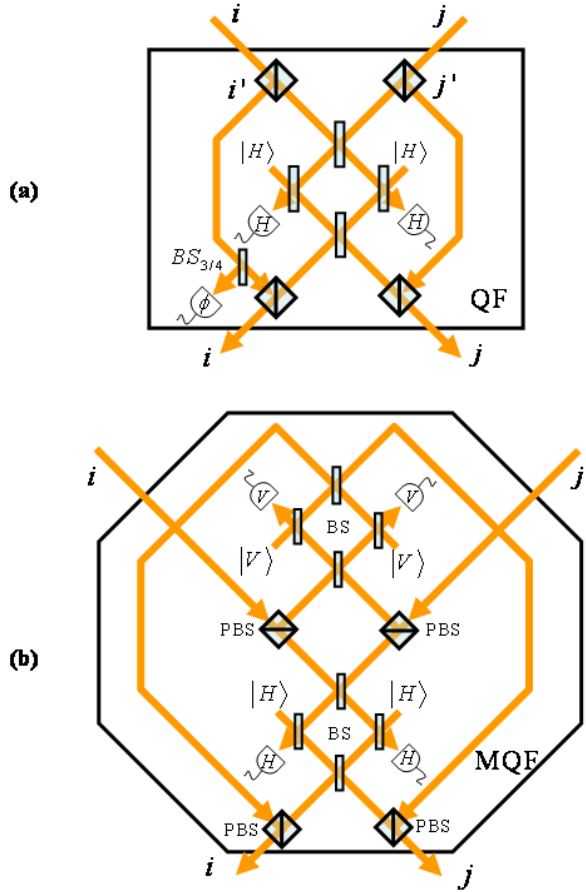} } \caption{
\label{fig:QF02} (color online) Implementation of (a) an original
and (b) a modified quantum filter}
\end{figure}
\end{center}
The original qubit QF \cite{Holger} depicted in Figure
\ref{fig:QF02} is useful for generating multi-qubit entangled
states deterministically. Two photons in modes $i$ and $j$ are
injected into the QF; successful detection occurs if and only if
their polarizations are the same. The most interesting feature is
that the injected photons  survive in the output modes after the
QF. Consider, as shown in Figure \ref{fig:QF02} (a), when two
input Bell-pairs (as in Eq. (\ref{six:eq04})) are passed through a
QF set up in mode 2 and 3. (The ${\rm BS}_{3/4}$ is required to
balance coefficients in the two different polarization modes). The
output state is simply the fusion of the initial Bell states,
i.e. the GHZ state $(|HHHH\>_{1234}+|VVVV\>_{1234})/\sqrt{2}$.

In cases where only a single photon is input there is a
fundamental asymmetry between horizontal and vertical input
photons in the workings of the QF. For example, the QF is
activated (a success detection is achieved) when a vertical
photon in mode $i$ is input  along with vacuum in mode $j$. Taking
into account vacuum inputs, the action of the QF circuit when
successful can be represented by a measurement operator
\begin{eqnarray}
\label{QF03}\hat{S}_{ij} &=& {1 \over 4} ( | H H \rangle_{ij}
\langle H H | + | V V \rangle_{ij} \langle V V |) + {1 \over 4} |
\o V \rangle_{ij} \langle \o V | \nonumber \\ &&+ {1 \over 2} (| V
\o \rangle_{ij} \langle V \o | + | \o \, \o \rangle_{ij} \langle
\o \, \o | ).
\end{eqnarray}
This asymmetry between $H,V$ is highly detrimental to our
purposes, as is the fact that the filter will trigger success
when only single photon is input.

To obtain a complete quantum filter for both polarizations, a
modification is required in the vertical part of the original
filter, and this takes two additional ancilla vertical photons.
This modified quantum filter (MQF) is shown in Figure
\ref{fig:QF02} (b). The  measurement operator for the MQF is
\begin{eqnarray}
\label{QF04}\hat{S'}_{ij} &=& {1 \over 8} ( | H H \rangle_{ij}
\langle H H | +| V V \rangle_{ij} \langle V V | ) \nonumber \\
&&+ {1 \over 4} | \o \, \o \rangle_{ij} \langle \o \, \o |).
\end{eqnarray}
Note that the ${\rm BS}_{3/4}$ with a vacuum input (which
essentially just dumped unwanted amplitude) is no longer needed
because the MQF balances the outcomes for both polarizations
naturally. The success probability of the MQF is 1/128.

\section{Implementation of a Bell pair from four single photons}
\label{Append03}
\begin{center}
\begin{figure}[t]
\resizebox{!}{7cm} {\includegraphics{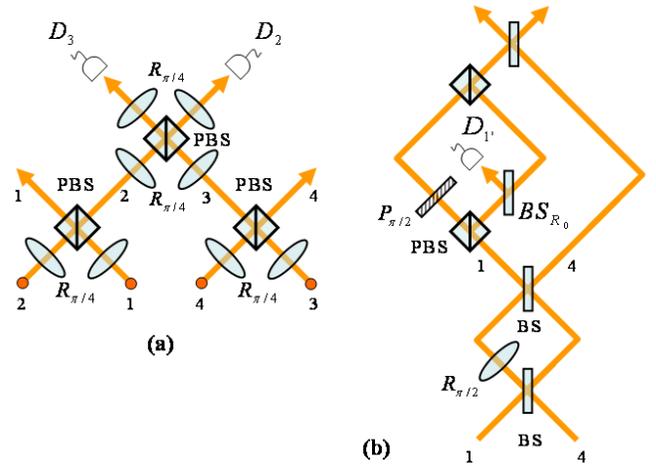} } \caption{
\label{NewS01} (color online) (a) The circuit $B$ creates a Bell
pair from four single photons in horizontal polarization when
single photons are detected at $D_2$ and $D_3$ and (b) An extra
circuit is required for correcting the case of two-photon
detection in circuit $B$.}
\end{figure}
\end{center}
We present a scheme to build a two-photon Bell state from four
single photons. Initially, all photons are prepared with
horizontal polarization. As shown in Figure \ref{NewS01}(a), each
photon passes through a $R_{\pi/4}$ in each mode and two PBSs are
applied in modes 1 and 2 as well as 3 and 4. Then, we use a
Type-II fusion gate between modes 2 and 3. If two different
polarized photons are detected in any mode, we obtain a typical
form of a Bell pair with a success probability 3/16, as noted in
\cite{Browne05,pan}. However, previously the detection of two
photons of the same polarization was regarded as a failure case.

In fact, we can transform this error state to a typical Bell pair
probabilistically with a simple circuit in Figure \ref{NewS01}
(b). For example, we assume that two horizontal photons are
detected in mode 2 and two photons remain in modes 1 and 4. The
outcome state contains a mixture of two-photon states in a single
spatial mode and a state of one photon in each mode. The first
part of the correction circuit consists of two BSs and a
$R_{\pi/4}$ (see the bottom of Figure \ref{NewS01} (b)). After
the first part, the total state is a bunching state made by two
equivalently polarized photons with unbalanced coefficients in
each mode, such as:
\begin{eqnarray}
\label{C3eq18} {1 \over 2\sqrt{3}} (|H^2 \rangle_1 - 3 |V^2
\rangle_1 + |H^2 \rangle_4 + |V^2 \rangle_4).
\end{eqnarray}

In the remainder part of the circuit, the $BS_{R_0}$ ($R_0=2/3)$)
with a vacuum input plays a key role to balance coefficients in
the final state and $P_{\pi/2}$ is a $\pi/2$ phase shifter in
mode 1. For successful cases, no photon detection at $D_{1'}$
reduces a coefficient of one photonic state in a certain spatial
mode. Finally, the total state is equal to
$|\phi^{-}\rangle_{1\,4}$ with a success probability 1/16.
Therefore, the optimal success probability is 1/4 to create a
Bell pair linear optically from four single photons.

\end{document}